\documentclass[letter]{aa}
\usepackage[varg]{txfonts}

\usepackage{epsfig}
\usepackage{epstopdf}

\usepackage{amssymb}

\begin{document}

\title{Unidentified  infrared bands do not correlate with C/O ratio in planetary nebulae}

\titlerunning{Unidentified  infrared bands and C/O ratio} 

\author{Fr\'ed\'eric Zagury\inst{1}
  \thanks{\emph{Present address:} Fondation Louis de Broglie,
  23 rue Marsoulan,
75012 Paris, France} 
    } 

\offprints{F. Zagury, \email{fzagury@wanadoo.fr}}
\institute{Fondation Louis de Broglie,
23 rue Marsoulan,
75012 Paris, France} 

\date{}

\abstract{
The concrete evidence adduced to support the widely held idea that unidentified  infrared bands (UIBs)  are enhanced in carbon-rich planetary nebulae (PNe) is a remarkable UIB 7.7~$\mu$m versus C/O ratio correlation plot for six PNe, obtained from air-born observations and published  in   1986 by  M. Cohen and coworkers.
However, the space-born data  presented by Cohen \& Barlow in 2005 undercut this correlation, and I show that  the larger dataset  they provide disproves a specific link between UIBs and carbon abundance in PNe. It also follows from these data  that interstellar UIB carriers  cannot originate from  the atmosphere of carbon-rich PNe.
}
\maketitle 
%\newpage
%%%%%%%%%%%%%%%%%%%%%%%%%%%%%%%%%%%%%%%%%%%%%%%%%%%%%  
\section{Introduction} \label{}
\begin{figure*}[]
\resizebox{2\columnwidth}{!}{\includegraphics{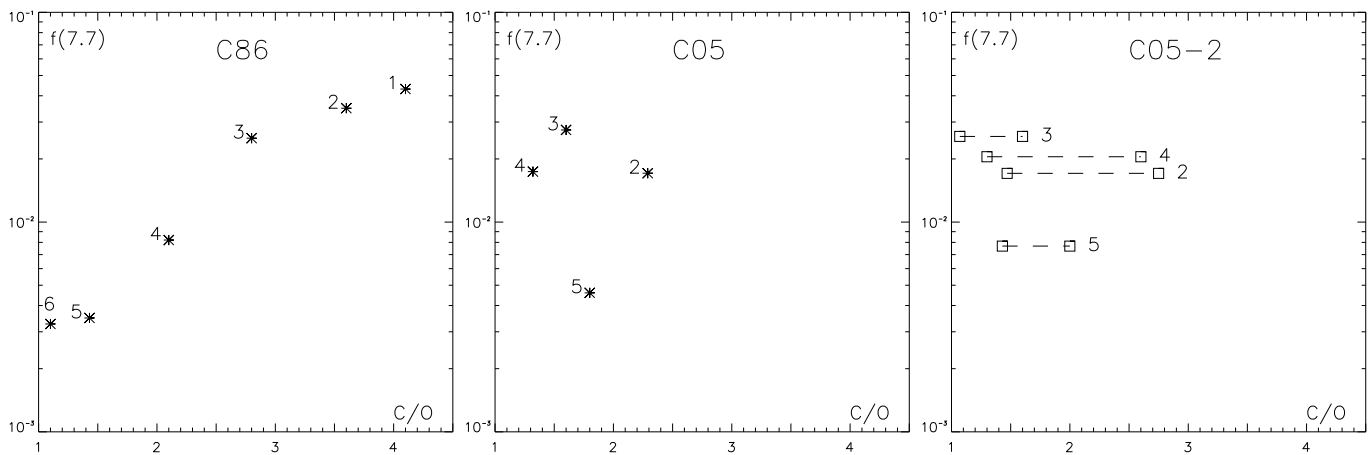}} 
\caption{Left plot: C05's Figure~11. Key: 1. J900; 2. NGC7027; 3. BD+30~3639; 4. IC5117; 5. IC418; 6. NGC6572.  
Middle: Same plot revised with C05's ISO data.
Right: $f(7.7)$ deduced from  Table~\ref{tbl:t2} and C/O range of values from Table~\ref{tbl:t1}.
} 
\label{fig:fig1}
\end{figure*}

%%%%%%%%%%%%%%%%%%%%%%%%%%%%%%%%%%%%%%%%%%%%%
The \citet[][hereafter, C86]{cohen86} finding of a correlation between the unidentified infrared bands (UIBs) and carbon abundance in planetary nebulae (PNe) (their Figure~11, reproduced in Figure~\ref{fig:fig1}) gave direct observational evidence of the organic nature of UIB carriers as propounded by   \citet[][]{leger84}.
This finding provided grounds for thinking that the carriers were synthesized in carbon-rich PNe prior to being released and disseminated in the interstellar medium. 
The correlation was updated  by \citet[][C05]{cohen05} from ISO (Infrared Space Observatory) spectra.
A correlation between UIBs and C/O ratio, however, conflicts with  the noticeable absence of UIBs in carbon stars \citep[][]{li20} and with the finding of  intense UIBs in  oxygen-rich PNe  \citep[][]{delgado14}.
 
Section~\ref{c86} of this paper shows that  C86's striking  Figure~11 is an artefact of unreliable pre-ISO spectra and arbitrarily selected C/O data.
The figure is not confirmed by C05's revised data.
The C05 paper itself uses a larger sample of PNe but fails to prove UIBs' dependency on C/O (Section~\ref{c05}).
%In fact, large $f(7.7)$ can be found in oxygen-rich circumstellar environments, such as HD100546's, as well as  in C-rich ones (Section~\ref{hd}).
Section~\ref{dis} concludes that there is no quantitative evidence linking UIBs to carbon-rich PNe.
Observations indicate rather  that UIBs are not related to carbon-rich environments, and that interstellar UIB carrier(s) do not originate from carbon-rich PNe.
%%%%%%%%%%%%%%%%%%%%%%%%%%%%%%%%%%%%%%%%%%%%%%%%%%%%%  
%%%%%%%%%%%%%%%%%%%%%%%%%%%%%%%%%%%%%%%%%%%%%%%%%%%%%  
%%%%%%%%%%%%%%%%%%%%%%%%%%%%%%%%%%%%%%%%%%%%%%%%%%%%%  
\begin{table*}
\begin{center}
\caption{Data for Figure~\ref{fig:fig1}.
\label{tbl:t1}
}		
\begin{tabular}{l|rr|rr|rr|rrr}
PN$^{(1)}$&  \multicolumn{2}{c|}{$I_{7.7}$$^{(2)}$}  &   \multicolumn{2}{c|}{$I_T$$^{(3)}$}   &   \multicolumn{2}{c|}{$f(7.7)$} & \multicolumn{3}{c}{C/O}\\
\hline
&  &      & &    &  &      & &  &  \\ 
& C86 & C05& C86$^{(4)}$&   C05  & C86$^{(4)}$&   C05 &  C86&   C05 &recent$^{(5)}$\\ 
 & \multicolumn{2}{c|}{\scriptsize{($10^{-18}$ W/cm$^2$)}} & \multicolumn{2}{c|}{\scriptsize{($10^{-15}$ W/cm$^2$)}}  &   \multicolumn{2}{c|}{} &  \multicolumn{3}{c}{}\\
&  &      & &    &  &      & &  &  \\  
NGC7027& $520.0$ & $350.0$ & $15.0$ &20.5 &    0.0349& 0.0171 & 3.6 &   2.29 &  1.5--2.7 \\ 
 BD+30$^o$3639 & $120.0$ & $135.0$  & $4.8$ &4.90 &   0.0251& 0.0275 & 2.8 &  1.60 & 1.1--1.6   \\ 
 IC5117 & $9.8$ &  $14.4$& $1.2$  &0.83 & 0.0082 &0.0174 & 2.1 &  1.32 &  1.3--1.4  \\ 
IC418 &  $14.0$ & $10.9$  & $4.0$ & 2.38 &  0.0035& 0.0046& 1.43 &  1.43 &  1.4--2.0  \\ 
\end{tabular}
\end{center} 
\begin{list}{}{}
\item[$(1)$] J900, not observed by ISO, is not part of C05's sample. 
NGC6572 was also discarded because UIBs do not emerge from its noisy sws-31901125 spectrum.
\item[$(2)$] Integrated intensity in UIB~7.7, background subtracted, from C86 (KAO spectra), C05 (ISO spectra). 
\item[$(3)$] Total estimated integrated  infrared flux.
\item[$(4)$] $f(7.7)$ and $I_T$ were calculated from C86's Table~2 and Figure~11.
\item[$(5)$] Range of PNe's nebular C/O values from recent literature.\\% \citep[see][]{pottasch10}.
 NGC7027: \citet{zhang05}. %bernard01, 
 BD+30 3639: \citet{bernard03}. %aller95,
IC5117:  \citet{hyung01}.  
IC418: \citet{delgado14}. 
\end{list}
\end{table*}
%%%%%%%%%%%%%%%%%%%%%%%%%%%%%%%%%%%%%%%%%%%%%%%%%%%%%  
%%%%%%%%%%%%%%%%%%%%%%%%%%%%%%%%%%%%%%%%%%%%%%%%%%%%%  
\begin{figure*}[]
\resizebox{2\columnwidth}{!}{\includegraphics{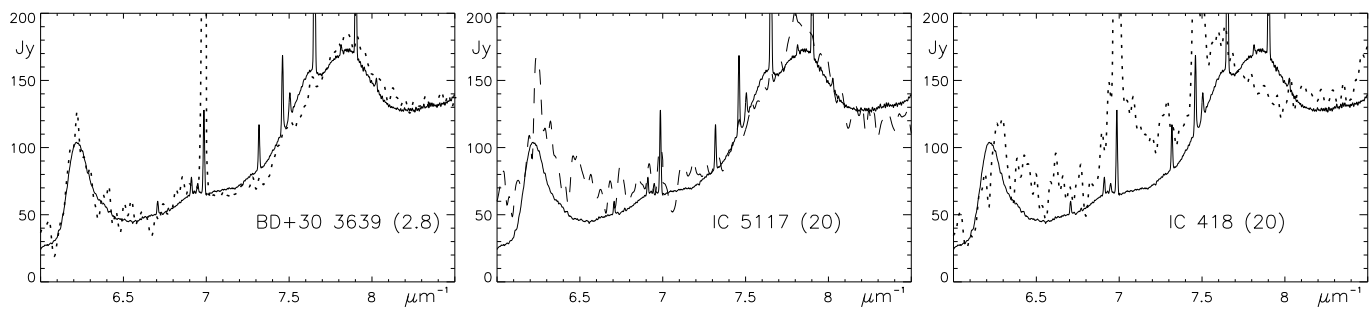}} 
\caption{ISO spectra of BD+30~3639, IC5117, IC418 (dotted lines) scaled  to match the spectrum of NGC7027.
Scaling factors  are in the parenthesis.
SWS spectra have been calibrated using the long wavelengths LWS  spectra (see C05 and Section~\ref{c86}).
Estimated calibration factors are 1, 1.15, 1, 1.2 for NGC7027,  BD+30~3639, IC5117, IC418.
} 
\label{fig:fig2}
\end{figure*}
%%%%%%%%%%%%%%%%%%%%%%%%%%%%%%%%%%%%%%%%%%%%%%%%%%%%%  
%%%%%%%%%%%%%%%%%%%%%%%%%%%%%%%%%%%%%%%%%%%%%%%%%%%%%  
\begin{table}
\begin{center}
\caption{C86 PNe's UIBs  relative to NGC7027's.
\label{tbl:t2}
}		
\begin{tabular}{l|rrr|rrrr}
PN&   \multicolumn{3}{c|}{$r_{uib}^{(1)}$}  &    \multicolumn{3}{c}{$f_r$$^{(2)}$} \\
\hline
&Fig.~\ref{fig:fig2}&C05&C86&Col. 2&C05&C86\\
\hline
NGC7027& $1$& $1$& $1$ & $1$ & $1$ &1 \\ 
 BD+30$^o$3639 & $0.36$& 0.38&0.23& $1.5$  & $1.6$ &0.72   \\ 
 IC5117 & $0.05$ &0.04&0.02&  $1.2$& $1.02$ & 0.23 \\ 
IC418 &  $0.05$ &0.03&0.027 &$0.45$  & $0.27$ & 0.10 \\
\end{tabular}
\end{center} 
\begin{list}{}{}
\item[$(1)$] Approximative ratio of UIBs (7.7 and 6.3) from the comparison of PNe's and NGC7027's ISO spectra (Figure~\ref{fig:fig2}), from C05, and from C86. 
\item[$(2)$] Ratio of $f(7.7)$ to NGC7027's deduced from column 2 ($I_T$ from C05), C05, C86.
\end{list}
\end{table}
%%%%%%%%%%%%%%%%%%%%%%%%%%%%%%%%%%%%%%%%%%%%%%%%%%%%%  

%%%%%%%%%%%%%%%%%%%%%%%%%%%%%%%%%%%%%%%%%%%%%%%%%%%%%  
\section{C86's Figure~11 in light of C05's updated data} \label{c86}
%%%%%%%%%%%%%%%%%%%%%%%%%%%%%%%%%%%%%%%%%%%%%
From a  sample of  six PNe, raised to 9 in \citet[][C89]{cohen89}, C86's Figure~11 (here Figure~\ref{fig:fig1}, left plot) demonstrates a  quasi-exponential growth of $f(7.7)$, the ratio of integrated intensity in the strongest  UIB (7.7~$\mu$m), $I_{7.7}$, to integrated infrared emission $I_T$, with  PNe's C/O ratio.
The figure  implies that carbon abundance in PNe determines the proportion of UIB carriers relative to dust and further suggests  that these carriers should be synthesized in carbon-rich PNe before their release in the interstellar medium.

C86's   $I_{7.7}$ relied on coarse observations (their Figure~1)  from  NASA's Kuiper Airborne Observatory (KAO).
Table~\ref{tbl:t1} shows that C05's ISO corrections to C86's $I_{7.7}$ amount to -30, +45, and -20\% for NGC7027, IC5117, IC418. 
Estimated from IRAS (7 to 140~$\mu$m) and ISO  (2 to 197~$\mu$m) respectively, $I_{T}$  in C86 and C05 disagree for most PNe, and are not even proportional:  +30\% for NGC7027 (from C86 to C05), -30\% for  IC5117, nearly -50\% for IC418  (Table~\ref{tbl:t1}).

Uncertainty on C05's  $I_T$  arises from the calibration correction applied to SWS spectra when they do not join smoothly with LWS spectra.
This rescaling process affects $I_{7.7}$ and to a lesser extent  $f(7.7)$ because uncertainties on $I_T$ and  $I_{7.7}$ partly cancel in  their ratio.
The other  uncertainty on $I_{7.7}$ lies in the   subtraction of the  continuum.
For NGC7027, \citet[][]{beintema96} find $I_{7.7}=1.8\,10^{-16}$ W/cm$^2$ (ISO data), against $3.5\,10^{-16}$ W/cm$^2$ in C05 (from the same data) and $5.2\,10^{-16}$ W/cm$^2$ in C86.
I tried to circumvent this difficulty by determining $I_{7.7}$ relative to a reference PN (NGC7027, Figure~\ref{fig:fig2} and Table~\ref{tbl:t2}),  since   C86's correlation relies on relative rather than absolute $I_{7.7}$  values.
The method confirms C05 but shows that infrared uncertainties remain large, especially when the shape of  UIBs varies (Figure~\ref{fig:fig2}, IC418).
Table~\ref{tbl:t2}  confirms that C86's results are incompatible with ISO observations.
The net  effects of ISO infrared corrections on C86's Figure~11 are to narrow  $f(7.7)$'s range of values and to switch the order of PNe on the $f(7.7)$ axis (Figure~\ref{fig:fig1}).
This  is already enough to ruin C86's $f(7.7)$ -- C/O correlation.

Determining abundances  in PNe is strongly model-dependent, one of several difficulties found in C/O estimates  \citep[see for instance][and Section~\ref{c05}]{stasinska02,delgado14}.
 C86 overestimated C/O for NGC7027, BD+30 3639, and IC5117.
For NGC7027, C86 took C/O = 3.6 from \citet{shields78} who warns that  his simplified model underestimates oxygen abundance and therefore overestimates C/O.
Other determinations available in 1986   are in better agreement with today's estimates (last column of Table~\ref{tbl:t1}).
\citet[][]{kaler81}, for instance, finds C/O  between  0.8 and 1.8 for most C86's PNe.
Shields'  value for NGC7027 was changed to 3.1 in C89 and 2.6 in C05;  BD+30$^o$3639's C/O = 2.8  was corrected to 1.6 in C89 and C05.

The middle  plot of Figure~\ref{fig:fig1} applies C05 updates to the C86 sample.
The right plot of the figure takes  $f(7.7)$ from  Table~\ref{tbl:t2}  with C05's value for NGC7027, and includes uncertainties on C/O.
It is clear from these plots that  C86's original correlation is an artefact of unreliable data.
%%%%%%%%%%%%%%%%%%%%%%%%%%%%%%%%%%%%%%%%%%%%%%%%%%%%%  
\section{C05} \label{c05}
%%%%%%%%%%%%%%%%%%%%%%%%%%%%%%%%%%%%%%%%%%%%%%%%%%%%%  
C89's Figure~20 ($f(7.7)$ versus C/O) rests on the same KAO observations as C86 and  suffers from the same uncertainties as C86's Figure~11. 
CPD-56$^o$8032, the most interesting  of the three PNe added to the sample, is given with $f(7.7)=0.08$ (the highest value  for C89's PNe) and C/O $=4.8$; these values are modified to $f(7.7)=0.027$ and  C/O $=13.1$ in C05.

Nevertheless, C05 resumed C86's problematic  with an enlarged dataset of 46 PNe observed by ISO, 17 of which only have detectable UIBs. 
For ten PNe with no or a noisy ISO LWS spectrum the authors  determined the far-infrared part of $I_T$ from IRAS  60/100~$\mu$m photometry and set the  relative error on $I_T$ (and $f(7.7)$) for these PNe at an additional  25\%.
Presumably, this uncertainty on $I_T$, which  was found to be even larger in Section~\ref{c86}, is a major  reason for M4-18's  outstanding $f(7.7)\simeq 0.05$.

CPD-56$^o$8032  (Hen 3-1333) and Hen 2-113 in C05 have abnormally large C/O, 13.1 and 10.4 respectively,   taken from \citet[][]{demarco97}.
These ratios  were obtained from collision excited lines (CELs) modeling which strongly depends on an adopted  electron temperature $T_e$.
For Hen 3-1333  \citet{danehkar21} found O/H = $3\,10^{-3}$, instead of $4.8\,10^{-4}$ in  De Marco et al., in agreement  with O/H $= 5\,10^{-3}$ for Hen 3-1333's spectral sibling  M4-18  \citep[][for M4-18 \citet{demarco99} found O/H  $=4.2\,10^{-4}$]{goodrich85}.
Neither De Marco et al. nor Danehkar were able to measure optical recombination lines (RLs)   for Hen 3-1333 and Hen 2-113, but the 18 other comparable PNe for which Danehkar could estimate C/O  from both CELs and RLs are all oxygen-rich (C/O << 1).
Uncertainties on C/O for Hen 3-1333, Hen 2-113 (and M4-18) are so large that, in the search for a $f(7.7)$--C/O correlation, the  PNe should be removed from the sample, as C05 actually did.
C/O error margin over the rest of the sample may still be large: NGC40, given as C-rich in C05, has C/O  between 0.17 and 1 according to  \citet[][]{delgado14}.  

\begin{figure}[]
\resizebox{1\columnwidth}{!}{\includegraphics{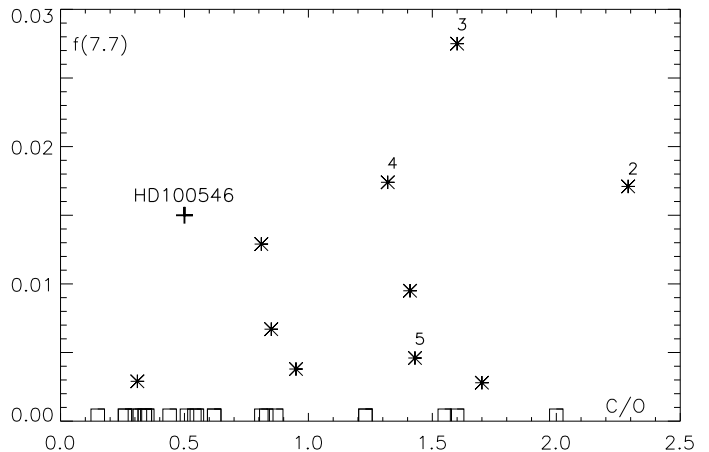}} 
\caption{C05's Figure~3, linear coordinates (see Section~\ref{c05}).
Asterix for C05's PNe with observed UIBs, squares for PNe  with UIBs not detected by ISO.
Numbers correspond to  Figure~\ref{fig:fig1}'s PNe.
} 
\label{fig:fig3}
\end{figure}

In Figure~\ref{fig:fig3} I have expunged C05's $f(7.7)$ versus C/O Figure~3 from the ten PNe with doubtful $I_T$, CPD-56$^o$8032, and Hen 2-113.
The plot is in linear instead of log-log coordinates --used by C05 because of the extreme values of Hen 3-1333, Hen 2-113, and M4-18--  for a better appreciation of the $f(7.7)$ versus C/O relationship.
PNe with ISO-undetected UIBs are  set to $f(7.7)=0$. 
Protoplanetary HD100546 \citep[][]{kama16,bouwman03}, 100~pc from the sun and  with solar-like C/O, was added to the plot.
The nebula's $f(7.7)$ can be estimated from its well-studied physical properties or from a comparison to NGC7027's  ISO spectrum and is in the range 0.015--0.022.
The plot would have benefited from oxygen rich PNe with intense UIBs observed by Spitzer \citep[such as NGC7026 and NGC6439, see][]{delgado14} had they been  observed by ISO.

C05's whole sample as well as  that of Delgado-Inglada \& Rodriguez (2014, see their Table 6) suggest that over two third of PNe are O-rich.
Their  C/O ratio barely exceeds 1.5, and PNe with weak or undetected UIBs (70\% of PNe) are found at all C/O.
PNe with UIBs are also observed at all C/O and $f(7.7)$ is generally under 0.02  (Figure~\ref{fig:fig3}).
This is less than in the regular (O-rich) interstellar medium.
I have  compared  ISO spectra of reflection nebula NGC7023 and of PN NGC7027, and found that $f(7.7)$ in NGC7023  is about three times that of NGC7027.
HI clouds in the Galactic plane also have $f(7.7)$ larger than 0.02 \citep[][and Section~\ref{dis}]{onaka00,sakon04}.
It is therefore very unlikely that PNe replenish the universe with UIB carriers and that the strength of PNe's UIBs depends  on their C/O ratio.

Figure~\ref{fig:fig3} obviously has nothing in common with C86's original finding (Figure~\ref{fig:fig1}, left plot) and disproves on a larger sample of PNe the strong correlation claimed in C86 and C89.
A systematic trend for $f(7.7)$ to increase with C/O, if any, would appear on the plot, which is not the case.
It is difficult to conclude from such a plot,  as Cohen \& Barlow did in resuming C86's ideas, that  $f(7.7)$ correlates with the C/O ratio of PNe.
%%%%%%%%%%%%%%%%%%%%%%%%%%%%%%%%%%%%%%%%%%%%%%%%%%%%%  
%%%%%%%%%%%%%%%%%%%%%%%%%%%%%%%%%%%%%%%%%%%%%%%%%%%%%  
%%%%%%%%%%%%%%%%%%%%%%%%%%%%%%%%%%%%%%%%%%%%%%%%%%%%  
\section{Discussion} \label{dis}
%%%%%%%%%%%%%%%%%%%%%%%%%%%%%%%%%%%%%%%%%%%%%
The C86--C89 finding of a strong dependency of $f(7.7)$ on C/O (Figure~\ref{fig:fig1}) was based on too few directions, questionable C/O values,  unreliable KAO infrared data, and a general absence of error estimates.
 \citet{cohen05} reconsidered the same $f(7.7)$ versus C/O problematic with new ISO data.
 Their data refute C86's Figure~11 and C89's Figure~20 correlations  and fail to provide evidence for a systematic growth of UIBs  with C/O.
Furthermore,  ISO data do not  demonstrate that the ratio of UIBs to dust thermal emission for PNe increases with their carbon abundance.
To date there is not a single indication of a direct link between  $f(7.7)$ and C/O in PNe,  consistent with the lack of UIBs in carbon stars and their observation in O-rich PNe.
Any new attempt to relate variations of $f(7.7)$ to C/O ratio in PNe would need to start from scratch, be more careful about error margins, and enlarge, for comparison, the data-set to include reflection nebulae and the regular interstellar medium.
I doubt  that such an undertaking would be successful.

Dust thermal emission in PNe peaks between 20 and 30~$\mu$m and represents most of the integrated intensity across both SWS and LWS ISO spectra.
In cooler media dust emission shifts towards the longer wavelengths and can be estimated from LWS spectra alone \citep[][]{onaka00}.
Figures~3 and 4 in \citet[][]{onaka00} show that in the $\rho$-Oph cloud (peak of thermal emission close to 100~$\mu$m) weakly illuminated by HD147889, $f(7.7)$  is  in the range 0.02-0.04,  which is much larger than generally observed  in  PNe (Figure~\ref{fig:fig3}).
C05's PNe observations may thus support Onaka et al.'s conclusion that $f(7.7)$ tends to diminish  in strongly ionized regions or under harsh radiation fields, which, again (see Section~\ref{c05}), casts doubts on the  possibility that C-rich PNe  are a privileged  site for synthesizing UIB carriers. 
 
The statement that UIBs were enhanced in carbon-rich PNe was a powerful support to the hypothesis that  UIB carriers had an organic molecular origin and were synthesized in  these PNe.
The opposite statement, that UIBs have no privileged link with C-rich PNe and cannot be related to their C/O ratio, is equally meaningful.
The implication is that  UIB carriers either are not synthesized in these PNe, or, that if they are,  they should as well be able to  form in situ wherever observed.
However, in the latter case their synthesis in (O-rich) low density HI cirrus devoid of molecular gas and unlikely to hide a complex chemistry would be difficult to justify.

\bibliographystyle{model3-num-names}
{}
%%%%%%%%%%%%%%%%%%%%%%%%%%%%%%%%%%%%%%%%%%%%%%%%%%%%%  
\end{document}